# USE OF DETAILED KINETIC MECHANISMS FOR THE PREDICTION OF AUTOIGNITIONS

F. Buda[1], P.A. Glaude[1], F. Battin-Leclerc[1*], R. Porter[2], K.J. Hughes[2] and J.F. Griffiths[2]

[1]DCPR-CNRS, 1, rue Grandville, BP 451- 54001 Nancy CEDEX - France

[2] School of Chemistry, University of Leeds, Leeds LS2 9JT - UK

## Abstract

This paper describes how automatically generated detailed kinetic mechanisms are obtained for the oxidation of alkanes and how these models could lead to a better understanding of autoignition and cool flame risks at elevated conditions. Examples of prediction of the occurrence of different autoignition phenomena, such as cool flames or two-stage ignitions are presented depending on the condition of pressure, temperature and mixture composition. Three compounds are treated, a light alkane, propane, and two heavier ones, n-heptane and n-decane.

## Introduction

Large-scale hydrocarbon (partial) oxidation processes form the basis of much of the (petro-) chemical process industry and in particular of the production of feedstock for

---

[*] Corresponding author:

fax: 00 33 3 83 37 81 20, e-mail: frederique.battin-leclerc@ensic.inpl-nancy.fr

polymer production (such as ethylene and propylene oxide, maleic and terephthalic acids). To prevent explosions, the European industry is in need of basic safety information, safety tools and predictive safety software, for instance, for design of equipment under all anticipated operating conditions. It is common knowledge that gas and vapour explosions are specific risks of the chemical industry. Much time and money are therefore spent in preventing such explosions, for example by eliminating all potential ignition sources. Commonly one thinks of the usual ignition sources such as flames, hot surfaces and sparks, but forgets that sufficiently high temperatures can lead to autoignition of the fuel-air-mixture.

An important feature of the oxidation of hydrocarbon is the possible occurrence of cool flame phenomenon at temperatures several hundred degrees below the minimum autoignition temperature. During a cool flame, or multiple cool flames, the temperature and the pressure increase strongly over a limited temperature range (typically up to 200°C), but the reaction stops before combustion is complete. Cool flames in themselves do not present a significant hazard, but they change the composition of the initial mixture, making it more reactive, with the accumulation of peroxide intermediates, which play an important part in causing normal ignition, as a two-stage process (Pekalski, Zevenbergen, Pasman, Lemkowitz, Dahoe, Scarlett, 2002).

A better understanding of autoignition and cool flame risks at elevated conditions and of the fundamentals of hydrocarbon oxidation leading to new and better tools for process engineering is therefore a very important challenge. The chemistry of hydrocarbon oxidation is very complex, as a mixture of hydrocarbons and air can react either in a slow combustion process or through an uncontrolled exponential increase in rate which leads to ignition, and can only be well described by using detailed kinetic mechanisms.

Detailed kinetic mechanisms are based on elementary steps, the rate constants of which depend only of temperature and pressure and can, therefore, be used in a predictive way. Detailed reaction mechanisms of the oxidation of hydrocarbons in the gas phase have been developed world-wide over several decades. Most studies have been confined to oxidation chemistry at temperatures above about 1000 K with more limited attention of detailed kinetic models to the important lower temperature regime. A major problem in constructing a detailed chemical kinetic model, especially at lower temperatures, is the very large number of possible reactions, products, and reaction intermediates involved. Because manual assembly of a comprehensive kinetic model is extremely difficult and prone to error, the only practical way to construct and use large models lies in the use of formal computer-based methods. The work presented here is based on the use of EXGAS, a fully automatic software able to generate detailed reaction mechanisms for gas-phase oxidation reactions (Warth, Battin-Leclerc, Glaude, Côme, Scacchi, 1998, 2000).

The purpose of this paper is to show how automatically generated detailed kinetic mechanisms can model the oxidation of alkanes and how they can predict the occurence of different autoignition phenomena, such as cool flames or two-stage ignitions, depending on the condition of pressure, temperature and mixture composition. We will first describe the case of a light alkane, propane, and then, that of two heavier compounds, n-heptane and n-decane.

# Automatically generated detailed mechanism of the oxidation of alkanes

Although the global reaction for the complete combustion of an alkane and its exothermicity, e.g. propane,

$$C_3H_8 + 5\,O_2 = 3\,CO_2 + 4\,H_2O, \quad \Delta_rH(1000K) = -2045.63 \text{ kJ/mole}$$

can be used to estimate the maximum explosion pressure, it is of no help to determine the temperature and the pressure where autoignition can take place, to explain the happening of cool flames (see further in the text) or to assess the by-products, such as carbon monoxide (CO) or aldehydes, which are also formed. For that is necessary to investigate in more details the reaction mechanism by considering the all possibilities of breaking and creation of chemical bonds, that is to say, by writing all the possible elementary reactions.

For the oxidation of alkanes containing at least three atoms of carbon, the possible elementary reactions of the reactant molecules and the derived radicals can be grouped according to the 7 following main types (taking as examples reactions related to the oxidation of propane):

1) Unimolecular and bimolecular initiation steps,

   e.g. $C_3H_8 \leftrightarrow \bullet CH_3 + \bullet C_2H_5$; $C_3H_8 + O_2 \leftrightarrow \bullet C_3H_7 + \bullet OOH$,

2) Reactions leading to alkenes from alkyl and hydroperoxyalkyl radicals:

   • Decompositions by beta-scission,

   e.g. $\bullet C_3H_7 \leftrightarrow \bullet CH_3 + C_2H_4$,

   • Oxidations to form the conjugated alkene and $\bullet OOH$,

e.g. •C$_3$H$_7$ + O$_2$ ↔ C$_3$H$_6$ + •OOH,

3) Additions of alkyl and hydroperoxyalkyl radicals on a molecule of oxygen,

e.g. •C$_3$H$_7$ + O$_2$ ↔ C$_3$H$_7$OO•, •C$_3$H$_6$OOH + O$_2$ ↔ •OOC$_3$H$_6$OOH,

4) Isomerizations of alkyl and peroxy radicals involving cyclic transition state,

e.g. CH$_3$CH$_2$CH$_2$OO• ↔ •CH$_2$CH$_2$CH$_2$OOH,

5) Decompositions of hydroperoxyalkyl and di-hydroperoxyal alkylradicals to form cyclic ethers, alkenes, aldehydes or ketones oxohydroperoxyalkanes, e.g. •CH$_2$CH$_2$CH$_2$OOH ↔ cyclic-C$_3$H$_6$O + •OH, e.g. CH$_2$OOHC•OOHCH$_3$ ↔ CH$_2$OOHC=OCH$_3$ + •OH,

6) Metathesis reactions to abstract an H atom from the initial reactant,

e.g. •OH + C$_3$H$_8$ ↔ H$_2$O + •C$_3$H$_7$,

7) Termination steps :
- Combinations of two free radicals,

  e.g. CH$_3$• + CH$_3$• ↔ C$_2$H$_6$,

- Disproportionations of peroxy radicals with •OOH,

  e.g. C$_3$H$_7$OO• + •OOH ↔ C$_3$H$_7$OOH + O$_2$,

As the rate constants of these elementary steps depend only on temperature and pressure, a model built in that way can be used to predict the autoignition conditions or the formation of by products. But to do so, this model has to be comprehensive, so that no important channel has been omitted. For that reason, a software EXGAS, has been developed in order to write systematically all the possible elementary reaction, by using an algorithm to ensure the comprehensiveness of the obtained mechanism (Warth et al., 2000). Details about the way the reactions of compounds containing less than two atoms of carbon and the decomposition of stable products obtained by the above-mentioned elementary steps are also taken into account can be found in previous papers

(Warth et al., 1998).

For a given alkane, or mixture of alkanes, EXGAS produces a file compatible with the CHEMKIN II softwares library (Kee, Rupley, Miller, (1993)) and containing all the relevant elementary reactions, as well as all the thermodynamic and kinetic data related (Warth et al., 1998). While the number of reactions involved in the mechanism can rapidly be important, the number of parameters needed to estimate the rate constants remains much more limited, as they are estimated using correlations mainly based on the type of elementary step and on the type of bond which is broken. For instance, the rate constant of the abstraction of an H-atom from the reactant (6) depends only on the type of H-atom which is abstracted, as shown in table 1. The same correlation is used for each linear alkane.

Figure 1 presents the global structure of a mechanism generated by EXGAS for an alkane (RH). Except at very high temperature (above 1500 K), the reaction starts by the reaction of alkane and oxygen molecules to give alkyl (•R) and hydroperoxy (•OOH) radicals. At low temperature (around 500-600 K), alkyl radicals react rapidly with oxygen molecules to give peroxyalkyl radicals (ROO•), which can by several reactions, as shown in figure 1, lead to the formation of peroxide species and of small radicals, which react with alkane molecules by metatheses (reaction 6) and regenerate alkyl radicals. The propagation of the reaction is a chain reaction, in which hydroxyl radicals (•OH) are the main chain carriers.

**FIGURE 1**

The two next parts show how models generated by EXGAS for the oxidation of propane, n-heptane and n-decane between 600 and 1500 K can reproduce observed experimental results.

# Prediction of the phenomena observed during the oxidation of propane in a closed vessel

The processes described above may be exemplified from the modelling of ignition in small vessels at sub-atmospheric pressures. A definitive pressure-temperature (p-$T_a$) ignition diagram for the combustion phenomena in propane was published by Newitt and Thornes (1937) and there have been many subsequent studies (Griffiths, Mohamed, 1996) but it is only very recently that the full richness of the behaviour has been modelled successfully (Griffiths, Hughes, Porter, 2004).

The simulation of multiple cool flames, and ignitions in a propane + oxygen mixture at a total pressure of 300 Torr (0.4 bar) is shown in Figure 2. The transition from cool flames through a three-stage ignition to two-stage ignition is brought about by a reduction in temperature from 610 K to 570 K and, as such, is a dramatic example of the consequence of the negative temperature dependence of reaction rate on temperature, as discussed below. The complete ignition diagram obtained by simulation is shown in Figure 3.

**FIGURES 2 AND 3**

# Modelling of the autoignition delay times of n-heptane and n-decane

The autoignition of n-heptane have been recently investigated in two complementary apparatuses : Minetti, Carlier, Ribaucour, Therssen, and Sochet (1995) has measured auto-ignition delays in a rapid compression for mixtures *n*-heptane/oxygen/argon/nitrogen for temperatures after compression from 600 to 900 K and pressures from 3.0 to 4.6 bar. Ciezki and Adomeit (1993) have studied the autoignition behind reflected shock wave of mixtures *n*-heptane/oxygen/argon for temperatures from 660 and to 1350 K and pressures from 3 to 42 bar.

Figure 4 shows the simulated temporal profiles of pressure and concentration of hydroxyl radicals vs. which can be obtained after the compression. This simulated profile of pressure exhibits the same behaviour as that of the experiments. After an induction period around 8 ms, a small rise of pressure is observed and is followed 25 ms later by a sharp increase corresponding to autoignition. That is characteristic of the two-stage autoignition, which starts first by a cool flame putted in evidence by the small pressure rise around 8 ms and is followed later by the actual autoignition. The time elapsed between the end of the compression and the sharp rise of pressure is the ignition delay time, which is a good characteristic of the reactivity of a compound.

As hydroxyl radicals are the main chain carriers, their concentration is a good indicator of the global rate of the reaction. Figure 4 presents the simulated profile of the mole fraction of these radicals and shows how a first acceleration of the reaction around 8 ms is followed by a slow down, while the second acceleration leads to autoignition.

Figure 4 shows also how during the cool flame an important concentration of hydrogen peroxide, which is a very explosive species used in rocket engines, is built on and is consumed only during the ignition itself.

**FIGURE 4**

Figure 5 presents a comparison between simulated and experimental ignition delay times related to both the shock tube and the rapid compression machine. It shows that a correct agreement is obtained. It is worth noting that our model is able to correctly predict both the position of the minima/maxima of the curve and the absolute value of auto-ignition delays, in a temperature region, which is known as difficult to model since there are important changes of reactivity. This figure clearly shows an important particularity of the oxidation of hydrocarbons, which is a zone of temperature, where the global rate of the reaction decreases with temperature, and which is usually called a "negative temperature coefficient" zone.

N-decane is the heaviest alkane for which autoignition data are available. Pfahl Fieweger and Adomeit (1996) have measured ignition delay times for n-decane/oxygen/nitrogen mixtures in a shock tube for temperatures between 660 and 1350 K and pressures of 12 and 50 bar. Experiments and simulations are presented in Figure 5. These results show that even at very different pressures (12 and 50 bars), our automatically generated mechanism is able to simulate the autoignition of n-decane in a shock tube in a satisfactory way.

Figures 5 and 6 shows that our model reproduce well the fact that the higher the pressure, the lower the ignition delay time. Modelling shows that an increase of the pressure from 50 to 100 bar induce only a 30% decrease of the ignition delay time.

**FIGURES 5 AND 6**

These figures illustrate also clearly the existence of a zone of negative temperature coefficient regime between 700 and 900 K, which is characteristic of the oxidation of alkanes and is displaced towards the higher temperatures when the pressure increases, and show that our detailed kinetic models are able to correctly capture these phenomenena.

Figure 7 displays a sensitivity analysis performed for n-heptane between 650 and 700 K, i.e. in the lowest part of the studied temperatures range. This figure shows that the primary reactions with the highest promoting effect are additions to oxygen molecule, which is the most sensitive reaction, metatheses and isomerizations. A secondary reaction, the decomposition of hydroperoxide species has also an important promoting effect. The primary reactions with the highest inhibiting effect are those directly competing with the additions to oxygen, i.e. the oxidations (to give alkenes and $HO_2\bullet$ radicals), the beta-scissions of alkyl radicals and the formations of cyclic ethers from hydroperoxyalkyl radicals.

**FIGURE 7**

The formation of hydroperoxides is extremely important, because they include a O-OH bond, which can easily be broken and lead to the formation of two radicals, which can in their turn react with alkane molecules to give alkyl radicals. This multiplication of the numbers of radicals in a chain reaction induces an exponential acceleration of reaction rates, leading in some conditions to spontaneous auto-ignition.

The reversibility of the addition of alkyl radicals to oxygen molecules when the temperature increases to the benefit of the formation of alkenes leads to an overall reduction of the reaction rate and induces the negative temperature dependent regime. When the temperature increase further, as a result of relatively weak exothermic propagation, other reactions (such as $H_2O_2 \rightarrow 2OH\bullet$ and $H\bullet + O_2 \rightarrow \bullet OH + \bullet O\bullet$) ensure the multiplication of the number of radicals and the reactivity accelerates again.

The decrease of reactivity with temperature explains the limited increase of temperature observed during a cool flame, since even a small elevation of temperature (a few ten of degrees) can lead to an important reduction of the reactivity, which, in some conditions almost stops the course of the reaction. In some cases, several accelerations/slow downs of the reaction can be observed, which are then observed as multiple cool flames.

# Conclusion

Detailed kinetic mechanisms, automatically generated based on generic types of elementary reactions, have been successfully used to model autoignition and cool flame characteristics for propane, n-heptane and n-decane. The experimental and modelled pressure-temperature ignition diagrams mapping the transition between the different observed oxidation phenomena (slow reaction, cool flames, multiple stages ignition) are qualitatively in good agreement for propane. The variations of autoignition delay times with temperature are well reproduced for n-heptane and n-decane, including the existence of a "negative temperature coefficient" zone, where the global reactivity of the mixture decreases when temperature increases. A sensitivity analysis has allowed the reactions of influence to explain these behaviours to be determined.

# Acknowledgement

This work has been supported by the European Commission within the "Safekinex" Project EVG1-CT-2002-00072.

# References

Ciezki, H.K., & Adomeit, G., (1993) Shock-tube investigation of self-ignition of n-heptane-air mixtures under engine relevant conditions, *Combustion and Flame*, 93, 421.

Kee, R.J., Rupley, F.M., & Miller, J.A. (1993) Chemkin II. A fortran chemical kinetics package for the analysis of gas-phase chemical kinetics, Sandia Laboratories Report, SAND 89 - 8009B.

Griffiths, J.F., & Mohamed, C., (1996), Experimental and numerical studies of oxidation chemistry and spontaneous ignition, Chapter 6 in Low-Temperature Combustion and Ignition (ed M.J. Pilling), Vol. 35, (p. 545) Amsterdam: Elsevier.

Griffiths, J.F., Hughes, K.J., & Porter, R. (2004) The role and rate of hydrogen peroxide decomposition during hydrocarbon two-stage autoignition, *Proceeding of the Combustion Institute*, 30, in press.

Minetti, R., Carlier, M., Ribaucour, M., Therssen, E., & Sochet, L.R. (1995) A rapid compression machine investigation of oxidation and autoignition of n-heptane : measurements and modeling, *Combustion and Flame*, 102, 298.

Newitt, D.M., & Thornes, L.S. , (1937) The oxidation of propane, *J. Chem. Soc.*, 1669

Pekalski A.A., Zevenbergen J.F., Pasman, H.J., Lemkowitz S.M., Dahoe A.E., & Scarlett B., (2002) The relation of cool flames and auto-ignition phenomena to process safety at elevated pressure and temperature, *Journal of Hazardous Material*, 93, 93.

Pfahl, U., Fieweger, K., & Adomeit, G. (1996) Shock tube investigation of ignition delay times of multi-component fuel/air mixtures under engine relevant conditions, Final Report, Subprogramme FK4, IDEA-EFFECT.

Warth, V., Stef, N., Glaude, P.A., Battin-Leclerc, F., Scacchi, G., & Côme G.M. (1998) Computed aided design of gas-phase oxidation mechanisms: Application to the modelling of normal-butane oxidation, *Combustion and Flame*, 114, 81.

Warth, V., Battin-Leclerc, F., Fournet, R., Glaude, P.A., Côme, G.M., & Scacchi G. (2000) Computer based generation of reactions mechanisms for gas-phase oxidation, *Computer and Chemistry*, 24 (5), 541.

# Figures captions:

**Figure 1** : Rules governing the generation of the primary mechanism for the oxidation of alkanes. Numbers refer to the types of elementary steps described in the text. Peroxide species are written in italics.

**Figure 2**: Simulated temperature – time profiles for multiple cool flames and two-stage ignition during propane combustion ($C_3H_8 / O_2 = 1$) in a closed vessel at a total pressure of 380 torr .

**Figure 3**: Simulated pressure - temperature ignition diagram for propane combustion ($C_3H_8 / O_2 = 1$) in a closed vessel. Solid line = ignition boundary, circles = the cool flame boundary, triangles = 3-stage ignition region, inverted triangles = 4-stage ignition region.

**Figure 4**: Simulated profiles of pressure and concentration of hydroxyl radicals after the compression in a rapid compression machine at 706 K and a pressure of 3.2 bar .

**Figure 5**: Modelling of ignition delay times obtained in a rapid compression machine for a pressure after compression between 3 and 4 bar (white triangles and thin line) and in a shock tube at 3.2 (black triangles and thin line), 13.5 (squares and thick line) and 42 (circles and broken line) bar for stoichiometric n-heptane/oxygen mixtures. Symbols are experimental results and lines refer to simulations.

**Figure 6**: Modelling of ignition delay times versus temperature for n-decane/air mixtures in a shock tube at 12 (triangles and thin line) and 50 (squares and thick line) bar and an equivalent ratio of 1.0 . Symbols are experimental results and lines refer to simulations.

**Figure 7**: Sensitivity analysis computed in the case of the autoignition of n-heptane at a temperature between 650 and 700 K and a pressure from 3.7 bar to 4.6 bar: the rate constant of each presented generic reaction has been divided by a factor 10. Only reactions for which a change above 5 % has been obtained are presented. To fit in the figure, the changes in % obtained for some very sensitive reactions have been divided by a factor 10 or 100.

.

**Table 1**: Examples of quantitative structure-reactivity relationships used for the estimation of rate constants in the case of the abstraction (reaction 6) of an H-atom from linear alkane (per H-atom abstractable). Rate constants are assumed to follow the modified Arrhenius law: $k = A\,T^b\,\exp(-E/RT)$, with the units cm$^3$, mol, s, kJ.

|  | Primary H-atom | | | Secondary H-atom | | |
| --- | --- | --- | --- | --- | --- | --- |
|  | *example for propane :* 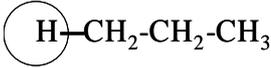 | | | *example for propane :* 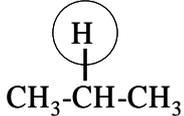 | | |
| Radicals | A | n | Ea | A | n | Ea |
| •O• | 13.23 | 0 | 32.8 | 13.11 | 0 | 21.7 |
| •H | 6.98 | 2 | 32.2 | 6.65 | 2 | 20.9 |
| •OH | 5.95 | 2 | 1.88 | 6.11 | 2 | -3.20 |
| •OOH | 11.30 | 0 | 71.1 | 11.30 | 0 | 64.8 |

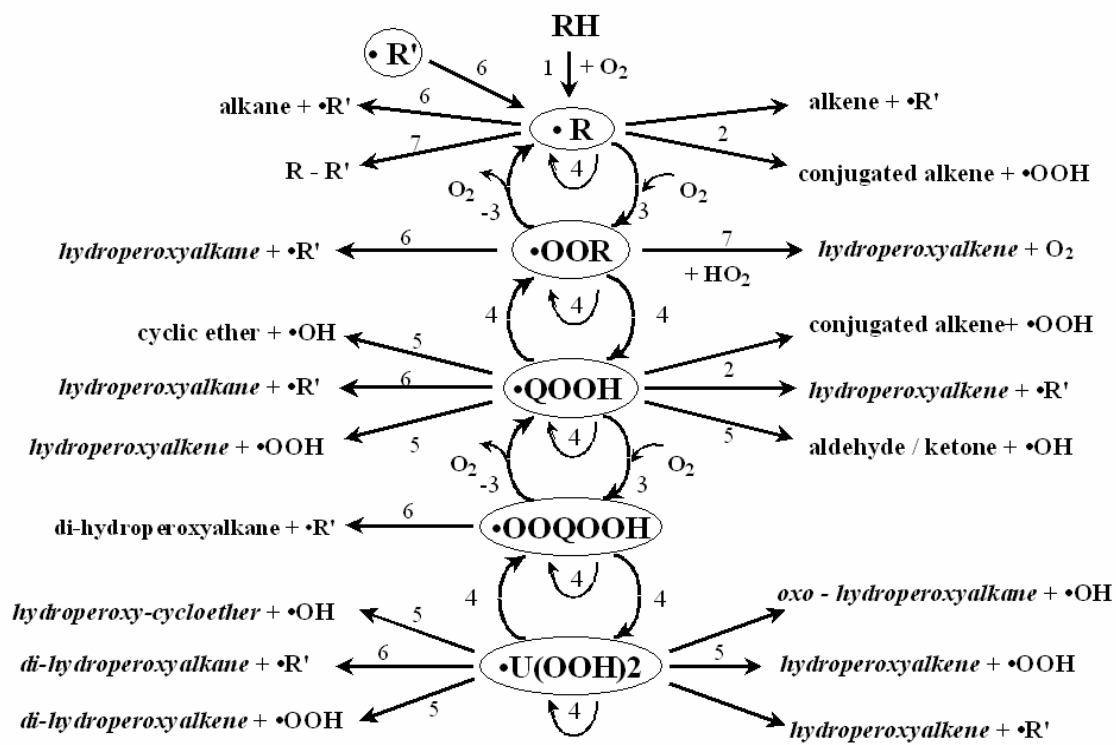

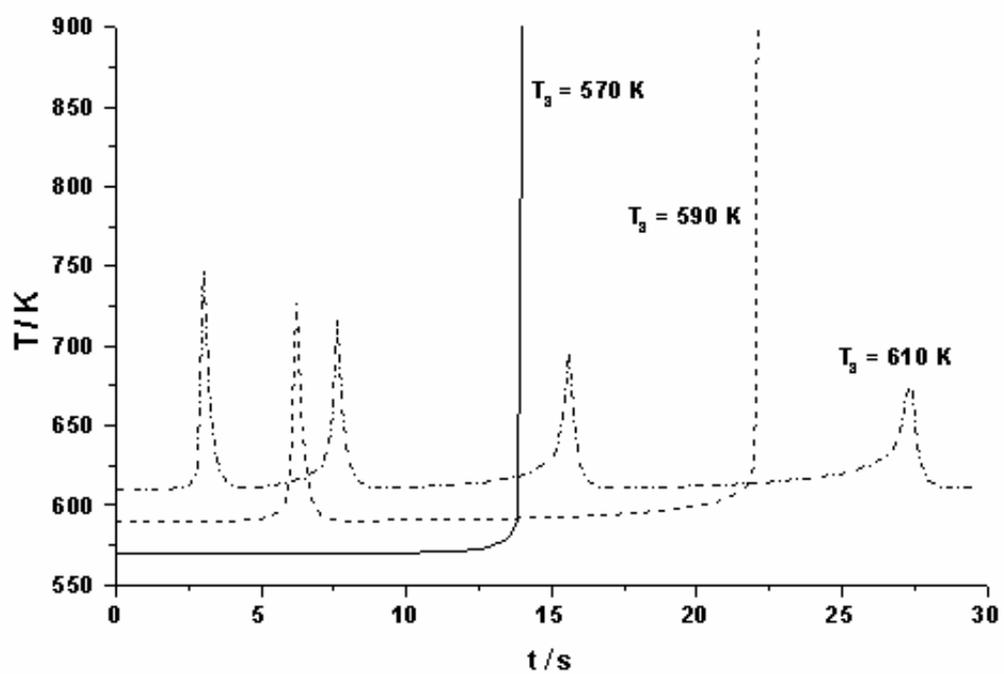

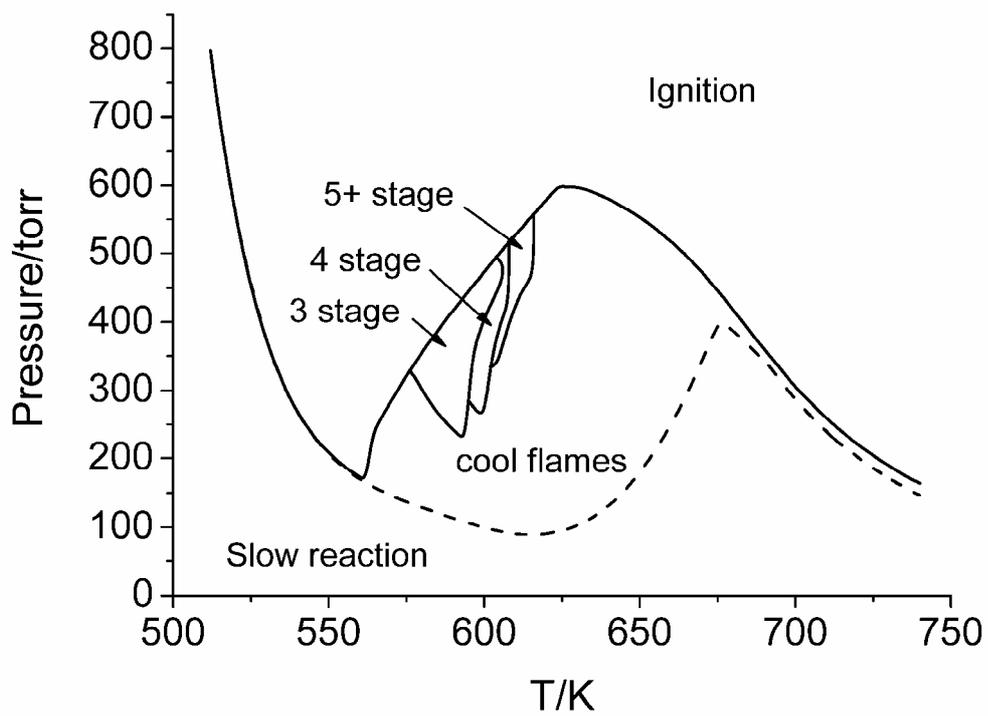

Figure 4

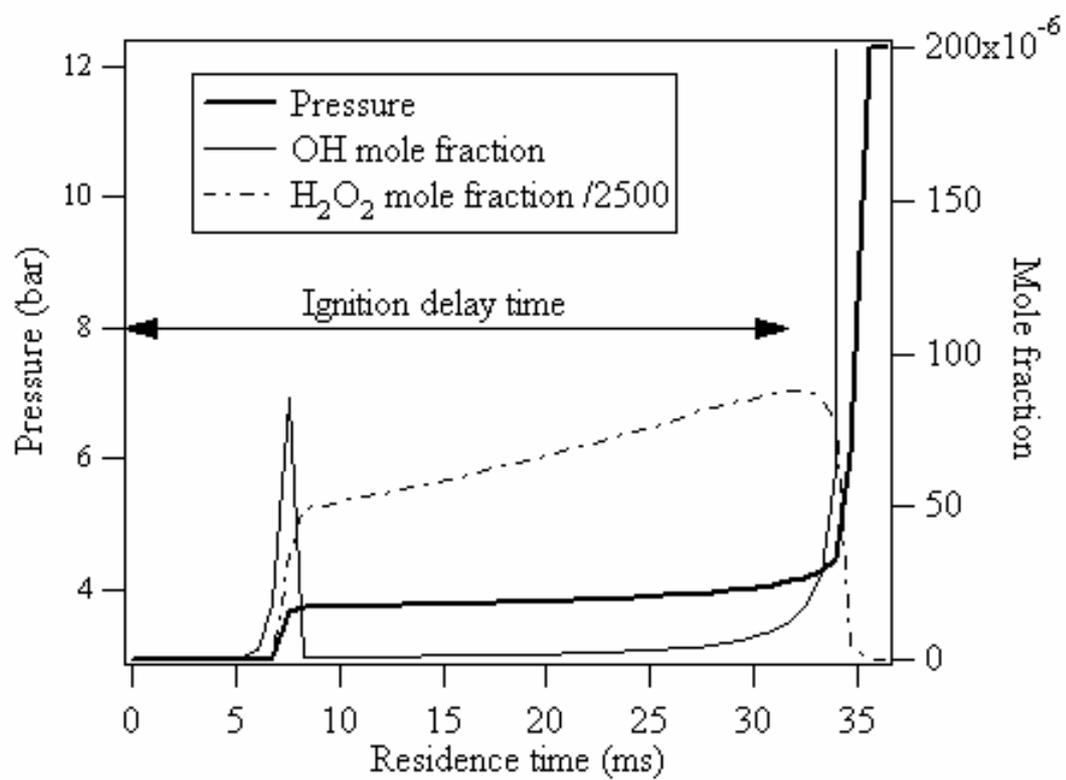

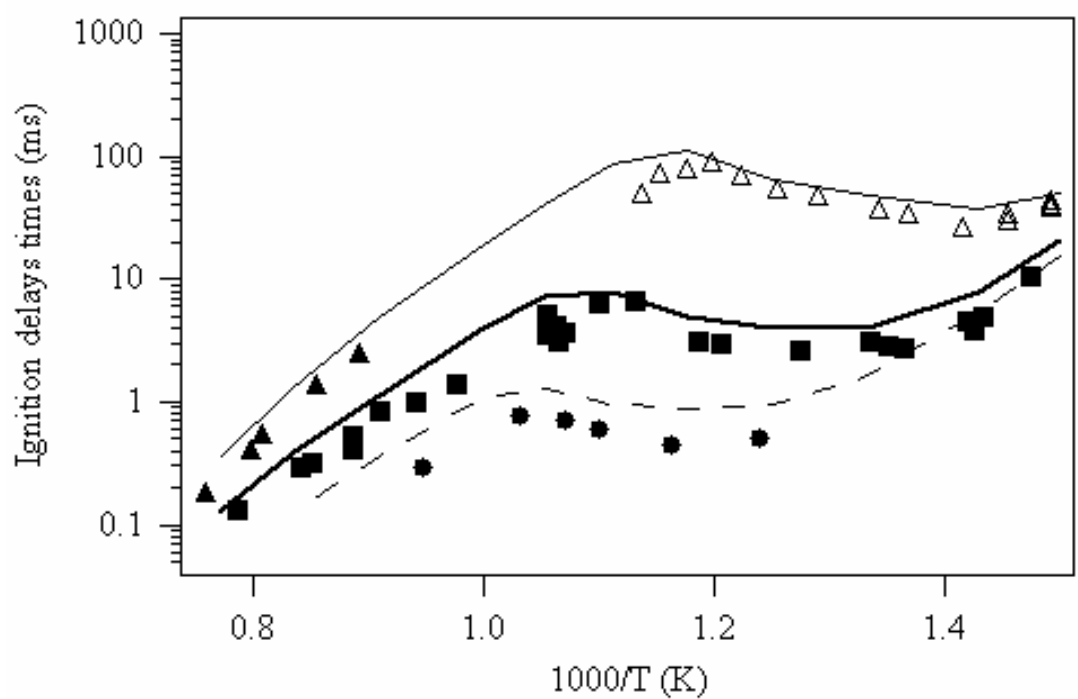

Figure 6

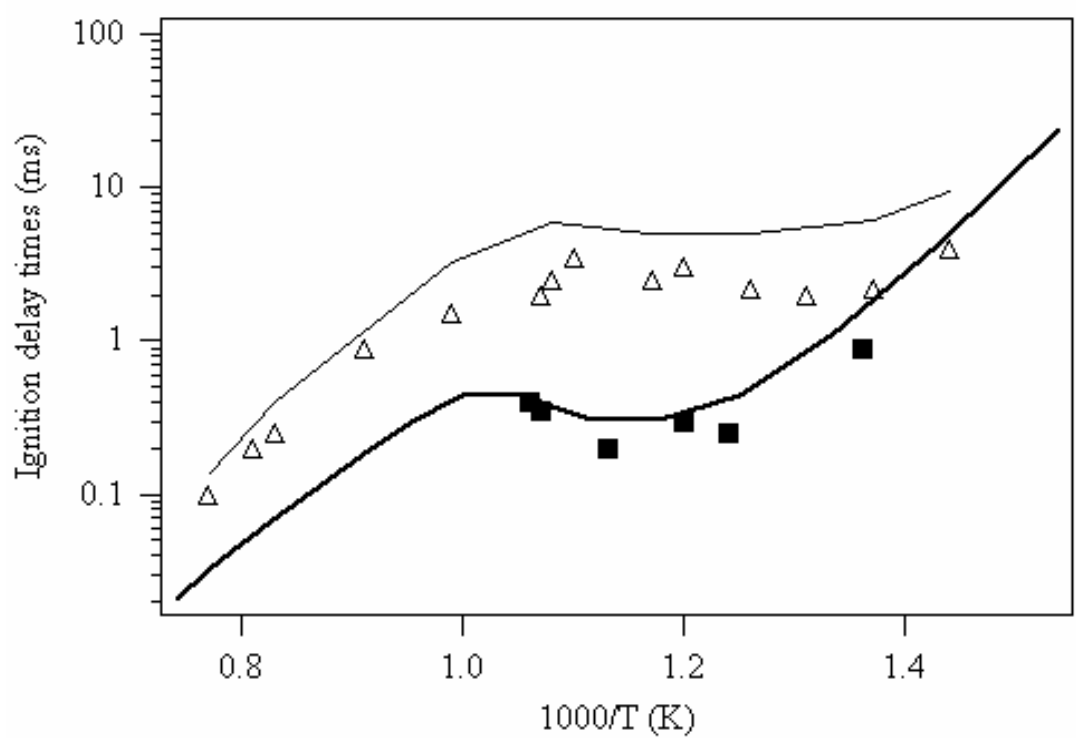

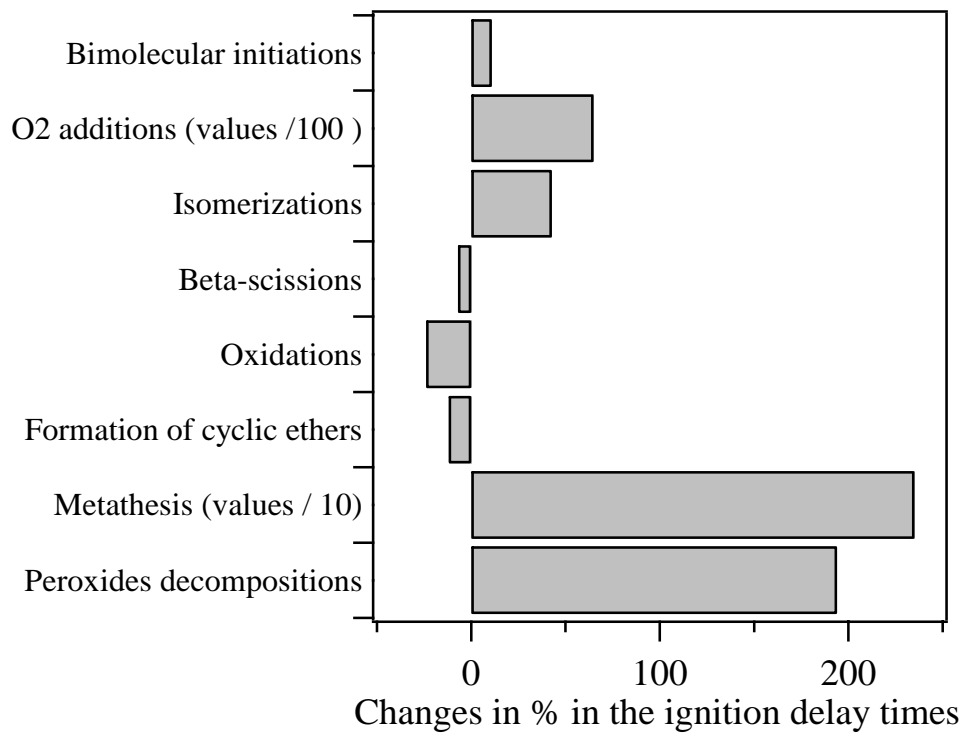